\documentclass[conference]{IEEEtran}
\IEEEoverridecommandlockouts
\usepackage{cite}
\usepackage{amsmath,amssymb,amsfonts}
\usepackage{algorithmic}
\usepackage{graphicx}
\usepackage{textcomp}
\usepackage{xcolor}
\usepackage{booktabs}
\usepackage{array}
\usepackage{hyperref}
\usepackage{times}
\usepackage{latexsym}
\usepackage{url}
\usepackage{tcolorbox}
\usepackage{colortbl}
\usepackage{amsmath}
\usepackage{subcaption}
\usepackage{multirow}
\usepackage{soul}
\definecolor{DarkGreen}{RGB}{1,130,5}
\def\BibTeX{{\rm B\kern-.05em{\sc i\kern-.025em b}\kern-.08em
    T\kern-.1667em\lower.7ex\hbox{E}\kern-.125emX}}
\begin{document}

\def\SR#1{{\color{red} {{#1}}}} % for Dr. Shafin

\title{LegalRAG: A Hybrid RAG System for Multilingual Legal Information Retrieval}

% \author{\IEEEauthorblockN{Muhammad Rafsan Kabir\textsuperscript{*}}
% \IEEEauthorblockA{\textit{Apurba-NSU R\&D Lab, ECE}\\
% \textit{North South University}\\
% Dhaka, Bangladesh\\
% muhammad.kabir@northsouth.edu}
% \and
% \IEEEauthorblockN{Rafeed Mohammad Sultan\textsuperscript{*}}
% \IEEEauthorblockA{\textit{Apurba-NSU R\&D Lab, ECE}\\
% \textit{North South University}\\
% Dhaka, Bangladesh\\
% rafeed.sultan@northsouth.edu}
% \and
% \IEEEauthorblockN{Fuad Rahman}
% \IEEEauthorblockA{\textit{Apurba Technologies}\\
% % \textit{name of organization (of Aff.)}\\
% Sunnyvale, CA 94085, USA \\
% fuad@apurbatech.com}
% \and
% \IEEEauthorblockN{Mohammad Ruhul Amin}
% \IEEEauthorblockA{\textit{Computer and Information Science}\\
% \textit{Fordham University}\\
% New York, USA\\
% mamin17@fordham.edu}
% \and
% \IEEEauthorblockN{Sifat Momen}
% \IEEEauthorblockA{\textit{Apurba-NSU R\&D Lab, ECE}\\
% \textit{North South University}\\
% Dhaka, Bangladesh\\
% sifat.momen@northsouth.edu}
% \and
% \IEEEauthorblockN{Nabeel Mohammed}
% \IEEEauthorblockA{\textit{Apurba-NSU R\&D Lab, ECE}\\
% \textit{North South University}\\
% Dhaka, Bangladesh\\
% nabeel.mohammed@northsouth.edu}
% \and
% \IEEEauthorblockN{Shafin Rahman}
% \IEEEauthorblockA{\textit{Apurba-NSU R\&D Lab, ECE}\\
% \textit{North South University}\\
% Dhaka, Bangladesh\\
% shafin.rahman@northsouth.edu}
% \thanks{\textsuperscript{*}Equal contribution}
% }

\author{\IEEEauthorblockN{Muhammad Rafsan Kabir\textsuperscript{1}\textsuperscript{*}, Rafeed Mohammad Sultan\textsuperscript{1}\textsuperscript{*}, Fuad Rahman\textsuperscript{2}, Mohammad Ruhul Amin\textsuperscript{3}, \\ Sifat Momen\textsuperscript{1}, Nabeel Mohammed\textsuperscript{1}, Shafin Rahman\textsuperscript{1}}

\IEEEauthorblockA{\textsuperscript{1}Apurba-NSU R\&D Lab, Department of Electrical and Computer Engineering, \\ North South University, Dhaka, Bangladesh}
\IEEEauthorblockA{\textsuperscript{2}Apurba Technologies, Sunnyvale, CA 94085, USA}
\IEEEauthorblockA{\textsuperscript{3}Fordham University, New York, USA}
\textsuperscript{1}\{muhammad.kabir, rafeed.sultan, sifat.momen, nabeel.mohammed, shafin.rahman\}@northsouth.edu \\
\textsuperscript{2}fuad@apurbatech.com, \textsuperscript{3}mamin17@fordham.edu\\
\textsuperscript{*}Equal Contribution
}

\maketitle

\begin{abstract}
Natural Language Processing (NLP) and computational linguistic techniques are increasingly being applied across various domains, yet their use in legal and regulatory tasks remains limited. To address this gap, we develop an efficient bilingual question-answering framework for regulatory documents, specifically the Bangladesh Police Gazettes, which contain both English and Bangla text. Our approach employs modern Retrieval Augmented Generation (RAG) pipelines to enhance information retrieval and response generation. In addition to conventional RAG pipelines, we propose an advanced RAG-based approach that improves retrieval performance, leading to more precise answers. This system enables efficient searching for specific government legal notices, making legal information more accessible. We evaluate both our proposed and conventional RAG systems on a diverse test set on Bangladesh Police Gazettes, demonstrating that our approach consistently outperforms existing methods across all evaluation metrics.
\end{abstract}


\section{Introduction}
In recent years, natural language processing (NLP) techniques have been widely applied in various domains, including healthcare, education, and governance, enabling the automated analysis and processing of large-scale textual data \cite{chowdhary2020natural}. One particularly significant area of research is the application of NLP techniques to government documents, specifically legal and regulatory texts such as government gazettes \cite{patel2012optical}. These documents contain critical legal information and government notices, but are often vast, unstructured, and linguistically diverse, making them both valuable and challenging for NLP applications. The effective application of NLP techniques to these documents is crucial to improve the accessibility of information and the policy-making process \cite{kawashima2024development}.

Although previous research has explored NLP applications in the processing of government regulatory documents, most efforts have focused on high-resource languages such as English, with techniques primarily used for document preprocessing \cite{10167495,10.5120/8794-2784}. In addition, some studies have examined financial regulations, such as those related to trading decisions. However, large language models (LLMs) in these methodologies often exhibit biases, and limited post-processing techniques have been used to address these issues \cite{zaremba2023chatgpt}. Despite some advancements, little effort has been made to extend such methodologies to low-resource languages like Bangla, where the lack of annotated datasets and specialized approaches presents significant challenges. This gap highlights the need for robust NLP solutions tailored to legal and regulatory documents in low-resource settings.

\begin{figure}[!t]
    \centering
    \includegraphics[width=\columnwidth]{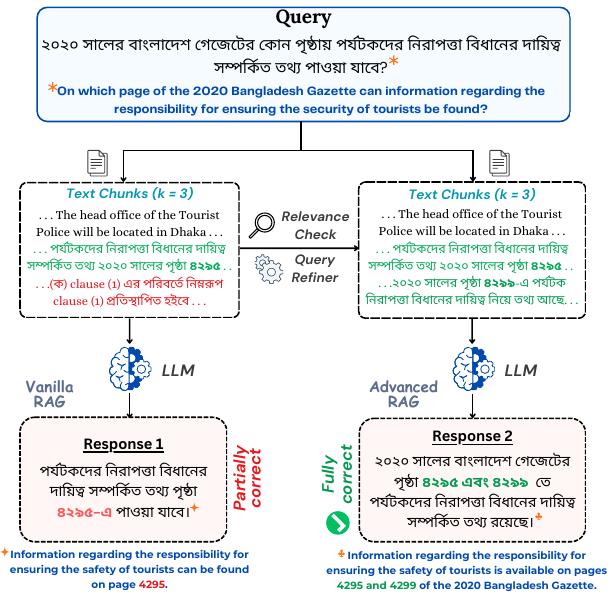}
    \caption{A sample response generated by the conventional RAG pipeline (Vanilla RAG) and our proposed pipeline (Advanced RAG) for a given user query. Our proposed advanced RAG pipeline improves the response by retrieving an additional \textcolor{DarkGreen}{relevant} text chunk for the LLM while eliminating an \textcolor{red}{irrelevant} one retrieved by the Vanilla RAG pipeline. The orange symbols indicate the English translation of the Bangla texts.}
    \label{fig1}
\end{figure}

To address these challenges, this work proposes a Retrieval-Augmented Generation (RAG) pipeline \cite{lewis2020retrieval} designed specifically for low-resource languages, focusing on Bangla, to process complex government regulatory documents. Given that Bangladeshi government gazettes, such as Bangladesh Police Gazettes, contain text in both Bangla and English, a multilingual RAG pipeline is essential to effectively handle the bilingual nature of these documents and retrieve context-specific information. As a first step, we implement a conventional multilingual RAG pipeline for a Question Answering (QA) task on Bangladesh Police Gazettes to facilitate the retrieval of precise legal notices and government information. However, because of the low resource nature, complexity, and unstructured format of these gazettes, the conventional RAG pipeline often struggles with retrieval accuracy and response generation. To overcome these limitations, we introduce a novel and enhanced multilingual RAG framework specifically designed for document-centric question-answering. Our proposed approach integrates an additional LLM that performs relevance checks on retrieved text fragments and refines the user query when necessary. This ensures that only the most relevant information is passed on to the final LLM responsible for generating responses, as shown in Figure \ref{fig1}. To evaluate the effectiveness of our approach, we curated a small yet diverse test set consisting of 168 question-answer pairs based on Bangladesh Police Gazettes. The performance of the system was assessed using two key criteria: human evaluation of generated responses and semantic similarity between generated outputs and ground-truth responses. Experimental results demonstrate that our proposed framework significantly outperforms the conventional RAG pipeline, highlighting its effectiveness in processing bilingual regulatory documents and improving QA accuracy. 

\noindent \textbf{Contributions:} To summarize, the major contributions of this work are as follows:

\begin{itemize}
    \item Introduction of an advanced multilingual RAG framework specialized for low-resource languages, specifically Bangla, to improve question-answering on complex bilingual legal documents.
    \item Unlike the conventional RAG pipeline, our approach incorporates an additional LLM for relevance checking and query refinement, ensuring that only the most relevant texts are passed to the generative model, thereby improving accuracy.
    \item A diverse Q\&A test set is curated based on Bangladesh Police Gazettes to assess the performance of the system. The evaluation, conducted using human judgment and semantic similarity metrics, demonstrates that the proposed framework outperforms the conventional RAG pipeline in handling bilingual regulatory texts.
\end{itemize}

\section{Related Works}
\noindent\textbf{NLP in Legal Documents:}
Natural language processing (NLP) has been widely employed in various critical domains, including the legal domain \cite{katz2023natural}. Zhong et al. \cite{zhong-etal-2020-nlp} provide an overview of legal artificial intelligence (LegalAI), specifically focusing on NLP techniques in the legal field. They highlight how NLP can be applied to legal tasks such as document analysis \cite{davie2021document} and information retrieval \cite{hambarde2023information}. The work by Chouhan and Gertz \cite{chouhan-gertz-2024-lexdrafter} contributes to legal NLP by drafting definitions for legislative documents using Retrieval-Augmented Generation (RAG) techniques \cite{lewis2020retrieval}. Their primary focus is on a collection of European (EU) documents. The LegalBench-RAG paper by Pipitone and Alami \cite{pipitone2024legalbench} introduces a RAG benchmark \cite{chen2024benchmarking} for the legal domain to evaluate the effectiveness of RAG-based systems. Despite these advancements in legal NLP, significant gaps still persist, especially in the domain of low-resource legal documents. Hence, our work contributes to this domain by primarily focusing on low-resource legal texts in Bangla.

\noindent\textbf{Retrieval Augmented Generation:} While pre-trained LLMs have demonstrated remarkable abilities in storing factual knowledge and achieving SOTA results across various NLP tasks, their capacity to effectively access and manipulate knowledge remains limited. Retrieval-Augmented Generation (RAG), introduced by Lewis et al. \cite{lewis2020retrieval}, enhances question-answering (QA) tasks by generating more specific, diverse, and factual responses. However, most RAG-based research has primarily focused on English applications, leaving a gap in studies addressing low-resource languages. Chirkova et al. \cite{chirkova2024retrieval} introduced the concept of multilingual RAG, exploring its effectiveness across different languages. Li et al. \cite{li-etal-2023-crosslingual} highlighted the transformative impact of recent LLMs, such as Llama \cite{dubey2024llama} and GPT \cite{achiam2023gpt}, on multilingual NLP. Their work proposed a cross-lingual retriever that maps Bangla text into a shared embedding space using cross-lingual prompts, demonstrating that high-resource language prompts can improve performance for low-resource languages. Based on these advancements, we employ advanced RAG techniques to develop an efficient QA system for Bangla legal documents.

\noindent\textbf{Advancements in RAG:}
Recent studies have highlighted significant advancements in RAG models, showcasing improvements in retrieval and generation performance. For instance, the Self-RAG framework introduced by Asai et al. \cite{asai2024selfrag} enhanced RAG by training a language model for both retrieval and generation. While promising, this method comes with computational challenges due to the need for model training. Yoran et al. \cite{yoran2024making} proposed a model that eliminates irrelevant contexts, thereby improving RAG's overall performance. In a similar vein, Koo et al. \cite{koo2024optimizing} introduced a query-document alignment score, which improved retrieval accuracy by 1.6\%. However, issues such as hallucinations and vague queries persist, highlighting the need for further improvements in query refinement. Sawarkar et al. \cite{sawarkar2024blended} proposed the Blended-RAG approach to optimize query processing. While some of these approaches focus on identifying irrelevant contexts, others aim to enhance QA performance through effective query refinement. Building on these advancements, our approach integrates a hybrid pipeline that combines a relevance check and a query refiner. This combination strengthens the system's ability to handle irrelevant contexts while simultaneously refining the queries for improved retrieval, resulting in a more robust and effective RAG-based solution.

\section{Methodology}

\subsection{Problem Formulation}
Given a collection of $n$ government gazettes, we concatenate them to form a single source document, denoted as $D$. This document is then divided into $m$ smaller text chunks, represented as $C = \{C_1, C_2, \ . \ . \ . \ C_m\}$, to facilitate efficient retrieval. Each chunk is processed through an embedding model, which generates corresponding embeddings $E = \{E_1, E_2 \ . \ . \ . \ E_m\}$. These embeddings are stored in a vector database to enable efficient similarity-based retrieval. During retrieval, relevant text chunks from $C$ are retrieved by computing the similarity between the stored embeddings $E$ and the embedding of the user query, denoted as $E_Q$. In the conventional vanilla RAG pipeline, the retrieved text chunks are directly passed to a generative LLM $L$, along with the query $Q$, to generate responses. However, in our proposed advanced RAG framework, the retrieved texts are first fed into a \emph{relevance check} and \emph{query refinement} LLM to ensure their relevance to the query $Q$. The refined and validated texts are then passed to the generative LLM $L$ for response generation.

\subsection{Data}
\noindent \textbf{Regulatory Document:}
In this study, we utilize a collection of official Bangladesh Police Gazettes published by the Bangladesh Government Press (BG Press). The Bangladesh Gazette is a weekly and regular publication by the Government of Bangladesh, with occasional additional issues. The Bangladesh Police Gazettes provide legal and regulatory updates on police operations, protocols, and administrative directives. For our experiments, we use all 13 available gazettes from the BG Press archive\footnote{\url{https://www.dpp.gov.bd/bgpress/index.php/document/gazettes/140}}, published between 2016 and 2023. These documents are a mixture of English and Bangla, a low-resource language, with the ratio shown in Figure \ref{doc}(B).
% These documents are predominantly written in Bangla, a low-resource language, with certain sections in English. 
The length of the gazettes varies, with the longest spanning 36 pages and the shortest consisting of a single page. The documents cover a wide range of topics, including updates to police regulations and descriptions of specialized police units such as the Anti-Terrorism Unit, Tourist Police, and River Police. Additionally, they contain administrative directives related to police training, investigation procedures, arrest protocols, and inter-agency coordination. The bilingual nature and complex structure of these gazettes necessitate the application of specialized NLP techniques to facilitate effective legal information retrieval. Table \ref{reg_doc} presents key statistics for the regulatory documents used in this study.

\begin{table}[!t]
    \centering
    \small
    \setlength{\tabcolsep}{1.5em}
    \caption{Key statistics from the employed Bangladesh Police Gazettes (2016–2023). }
    \label{reg_doc}
    \begin{tabular}{l r}
        \toprule
        \textbf{Metric} & \textbf{Value} \\
        \midrule
        Total Gazettes & 13 \\
        Total Pages & 81 \\
        Minimum Pages in a Gazette & 1 \\
        Maximum Pages in a Gazette & 36 \\
        Average Pages per Gazette & 6.23 \\
        \bottomrule
    \end{tabular}
\end{table}

\subsubsection*{Data Preprocessing} 
The Bangladesh Police Gazettes are originally available in PDF format. However, standard PDF loaders do not perform well on Bangla texts, necessitating their conversion into a text-based format. To achieve this, Optical Character Recognition (OCR) techniques were applied \cite{patel2012optical}. The preprocessing steps are outlined below:
\begin{enumerate}
    \item Each page of the PDF is first converted into an image using the \texttt{pypdfium2} library and saved in JPEG format for efficient processing.
    \item Tesseract OCR is configured to recognize both Bangla and English texts from the images. Several preprocessing techniques, including median filtering, contrast enhancement, and black \& white conversion, are applied to enhance the image clarity and improve OCR accuracy.
    \item The extracted text from each page is then combined into a structured text document, making it suitable for further analysis.
\end{enumerate}

\begin{figure}[!t]
    \centering
    \includegraphics[width=\columnwidth]{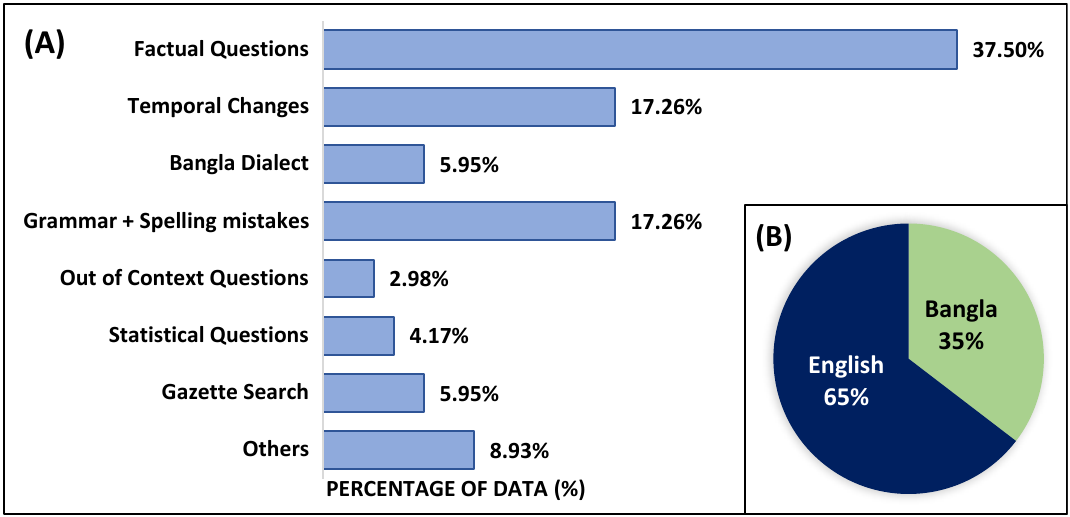}
    \caption{\textbf{(A)} Distribution of question-answer pair domains in the curated evaluation dataset. \textbf{(B)} Language distribution in the bilingual regulatory document (Bangladesh Police Gazettes).}
    \label{doc}
\end{figure}

\begin{figure*}[!t]
    \centering
    \includegraphics[width=\linewidth]{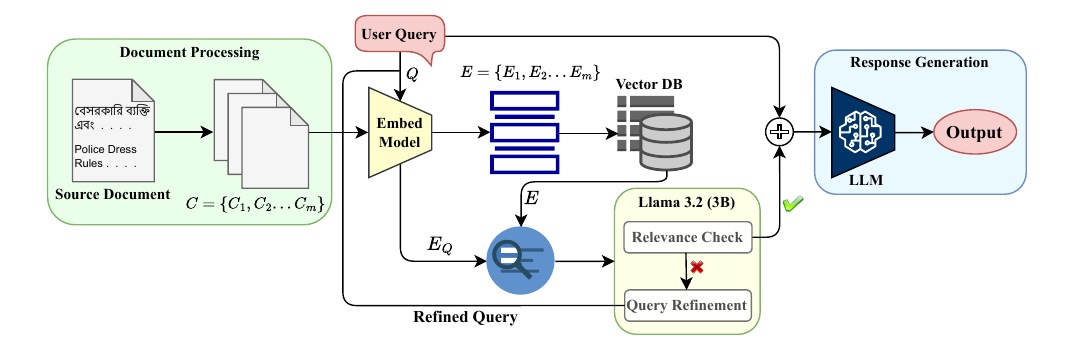}
    \caption{Proposed RAG pipeline for multilingual legal document question-answering. The yellow box highlights the \textbf{Relevance Check} and \textbf{Query Refinement} processes, which are introduced in our proposed novel framework to enhance the existing vanilla RAG pipeline.}
    \label{rag}
\end{figure*}

\noindent \textbf{Evaluation Dataset:}
The preprocessed regulatory document texts are segmented and provided as input to OpenAI's GPT-4o \cite{hurst2024gpt}, using tailored prompts to generate a series of question-answer pairs, both specifically in Bangla. Two of the authors serve as evaluators, cross-checking the generated pairs to eliminate inaccuracies. The evaluation set is carefully curated to ensure diversity, allowing for a rigorous assessment of the model’s ability to effectively comprehend and process a low-resource language. The dataset includes questions spanning multiple categories, such as factual inquiries, temporal changes, statistical reasoning, and variations in Bangla dialects, as depicted in Figure \ref{doc}(A). This diversity is crucial for testing the system's ability to handle various aspects of language comprehension and reasoning, from retrieving specific facts to interpreting context-dependent meanings in both formal and colloquial Bangla.

Additionally, out-of-context questions are incorporated to assess the system’s ability to filter relevant information and avoid distractions from unrelated content. The evaluation set also includes questions with spelling and grammatical errors to test the model’s robustness in handling common linguistic inaccuracies, reflecting real-world challenges with noisy or complex text. By drawing questions from diverse contexts, the dataset effectively evaluates both general language processing capabilities and the unique challenges associated with Bangla, a language with limited resources. In total, the curated dataset consists of 168 question-answer pairs, thoughtfully distributed across categories to ensure a balanced representation of diverse language processing tasks.

\subsection{Retrieval-Augmented Generation}
To develop an effective Q\&A system for legal documents that facilitates the searching and retrieval of specific government announcements and notices, we employ Retrieval-Augmented Generation (RAG) \cite{lewis2020retrieval}, as it has been shown to effectively reduce hallucinations in large language models (LLMs) \cite{shuster-etal-2021-retrieval-augmentation, ayala-bechard-2024-reducing}. In this study, we implement two types of RAG pipelines for legal documents: (a) a vanilla RAG pipeline and (b) an advanced RAG pipeline incorporating a relevance check and query refinement. The first approach is a basic multilingual RAG pipeline capable of processing both English and Bangla, as the source documents contain both languages. The second approach extends this capability by integrating an additional layer of relevance checking and query refinement, making it a unique and novel RAG pipeline. To generate responses, we utilized three popular generative LLMs: Mixtral 8×7B \cite{jiang2024mixtral}, Llama 3.1 (8B) \cite{dubey2024llama}, and Gemma 2 (9B) \cite{team2024gemma}.

\noindent \textbf{Vanilla RAG Pipeline for Legal Documents:}
To develop a Q\&A system for Bangladeshi regulatory documents, we implement a multilingual Vanilla RAG pipeline designed to process regulatory texts containing both English and Bangla. The steps involved in designing this pipeline are outlined below: \textbf{(1)} The regulatory documents, specifically the Bangladesh Police Gazettes, are loaded and preprocessed. To ensure efficiency in embedding generation, the documents are segmented into manageable text chunks $C = \{C_1, C_2, \ . \ . \ . \ C_m\}$ using the \texttt{RecursiveCharacterTextSplitter}. \textbf{(2)} A multilingual embedding model, \texttt{BAAI/bge-m3} \cite{bge-m3}, is employed to generate embeddings $E = \{E_1, E_2 \ . \ . \ . \ E_m\}$ for both Bangla and English texts. This model is selected for its ability to produce high-quality embeddings across multiple languages. \textbf{(3)} The generated embeddings $E$ are stored in ChromaDB, a vector database chosen for its efficiency in handling high-dimensional data, facilitating fast retrieval during the query process. \textbf{(4)} A retriever is initialized to fetch relevant text chunks based on user queries $Q$. Maximum Marginal Relevance (MMR) is utilized as the retrieval strategy to ensure that the retrieved results are both relevant and diverse. \textbf{(5)} When a user query $Q$ is submitted, its embeddings $E_Q$ are computed using the same embedding model. The retriever then measured the similarity between the query embeddings ($E_Q$) and the stored text embeddings ($E$) in the vector database, retrieving the $k$ most similar text chunks based on their similarity scores. \textbf{(6)} The retrieved text chunks are concatenated with the query and an appropriate instruction prompt before being passed to a generative LLM. The LLM then generates a response using the retrieved context, ensuring relevance and accuracy while minimizing hallucinations.

The performance of vanilla RAG is acceptable for various domains, such as science and finance. However, as shown in Table \ref{domain_com}, it exhibits significant limitations when applied to the legal domain. Legal documents, particularly low-resource gazettes, are often unstructured, contain complex terminologies, and are frequently multilingual. These challenges motivate us to explore an advanced RAG pipeline capable of effectively handling such complexities.

\begin{table}[!t]
    \centering
    \caption{Performance of Vanilla RAG across multiple domains, demonstrating poor performance in the legal domain and emphasizing the need for an improved RAG pipeline.}
    \label{domain_com}
    \begin{tabular}{l l c}
    \toprule
        \textbf{Domain} & \textbf{Dataset} & \textbf{Semantic Similarity}\\
    \midrule
        Finance & SEC 10-Q &  0.895\\
        Scientific & Llama-2 Paper &  0.905 \\
        Legal & Bangladesh Gazettes & 0.760\\
    \bottomrule
    \end{tabular}
\end{table}

\noindent \textbf{Advanced RAG Pipeline for Legal Documents:}
To further enhance the performance of the legal document Q\&A system, we develop a novel and advanced RAG pipeline that incorporates a relevance check and query refinement mechanism. This advanced pipeline follows a structure similar to the vanilla RAG pipeline but introduces an additional large language model (LLM) to perform relevance checking and query refinement, thereby improving the accuracy and effectiveness of the retrieval process.

Specifically, an LLM is integrated after the retriever to assess the relevance of the retrieved text chunks in relation to the user query $Q$. The retrieved texts, along with the query $Q$, are passed to this LLM, which determines whether the retrieved chunks are relevant. If they are deemed relevant, they are forwarded to the generation LLM for response generation. If they are not relevant, the LLM is instructed, through prompting, to refine the user query while preserving its original meaning. The refined query is then used to retrieve new text chunks, which undergo the same relevance-checking process. This refinement cycle is applied for a maximum of $n$ iterations (three in our case), ensuring an optimal balance between retrieval performance and computational efficiency. To ensure the overall RAG pipeline remains computationally efficient, we employ Llama 3.2 (3B), a multilingual LLM with only 3 billion parameters, for the relevance check and query refinement tasks. Figure \ref{rag} illustrates the proposed RAG pipeline, highlighting the relevance check and query refinement mechanisms.

\section{Experiments}
\subsection{Setup}
% \noindent\textbf{Dataset.}

\noindent\textbf{Implementation Details:} 
We develop a question-answering framework for low-resource legal documents, specifically the Bangladesh Police Gazettes. To process the data, we utilize \texttt{PyPDFium2} for converting PDF files into images, \texttt{Pillow} for image processing, and \texttt{PyTesseract} for Optical Character Recognition (OCR) \cite{4376991}. The preprocessed data serves as the source document for the question-answering system, which is built using a Retrieval-Augmented Generation (RAG) approach. We implement two distinct types of RAG pipelines: (a) a Vanilla RAG and (b) an Advanced RAG pipeline. In the proposed Advanced RAG pipeline, we incorporate Llama 3.2 (3B) \cite{dubey2024llama} for relevance checking and query refinement.
For response generation, we experiment with three large language models (LLMs): Llama 3.1 (8B) \cite{dubey2024llama}, Gemma 2 (9B) \cite{team2024gemma}, and Mixtral 8×7B \cite{jiang2024mixtral}. To ensure precise and factually accurate responses, we set the sampling temperature to 0.1, as legal question-answering tasks require exact information rather than creative responses. Due to the large parameter sizes of these models, we employ the Quantized Low-Rank Adapters (QLoRA) configuration \cite{dettmers2023qlora} to optimize model loading. All experiments are conducted using the \emph{PyTorch} framework, and the computations are performed on a single NVIDIA Tesla P100 GPU.

\begin{table}[t]
\centering
\caption{Demographic details of evaluators.}
\label{evaluator}
\begin{tabular}{@{}ll@{}}
\toprule
\textbf{Evaluator Criteria} & \textbf{Details} \\ \midrule
Total No. of Evaluators    & 3                    \\
No. of Female Evaluators   & 1                    \\
No. of Male Evaluators     & 2                    \\
Average Age                & 25                   \\
Proficiency in Bangla/English & Strong           \\
Prior Knowledge in Legal Documents & Yes        \\
Professional Background   & Undergrad Law Student, Master's \\
                          & Law Student, Legal Assistant            \\
\bottomrule
\end{tabular}
\label{tab:evaluator_profile}
\end{table}

\noindent\textbf{Evaluation:} 
To assess the performance of the proposed RAG-based question-answering (Q\&A) pipeline, we evaluate it on a diverse test set comprising 168 question-answer pairs from multiple domains. The evaluation is conducted using two distinct criteria: (a) semantic similarity between the system-generated responses and the ground-truth answers and (b) human evaluation of the generated responses.
For semantic similarity evaluation, we compute the cosine similarity between each generated response and its corresponding ground-truth answer. The overall mean (\( \mu \)) and standard deviation (\( \sigma \)) of the similarity scores are then calculated. The cosine similarity between two vectors \( \mathbf{u} \) and \( \mathbf{v} \) is given by:
\begin{equation}
    \text{Cosine Similarity} = \frac{\mathbf{u} \cdot \mathbf{v}}{\|\mathbf{u}\| \|\mathbf{v}\|}
\end{equation}
For human evaluation, three independent evaluators assess each generated response. All evaluators possess a strong understanding of legal government documents, particularly Bangladeshi gazettes. The evaluation team consists of two males and one female, with an average age of 25. All evaluators are native Bangladeshi speakers, making them well-suited for assessing responses related to Bangladesh Police Gazettes. They also demonstrate strong proficiency in both Bangla and English. Among them, one is an undergraduate student and another is a master's student, both from the Law department, while the third is a legal assistant. Table \ref{evaluator} presents some demographic details of the evaluators. 

\begin{figure}[!h]
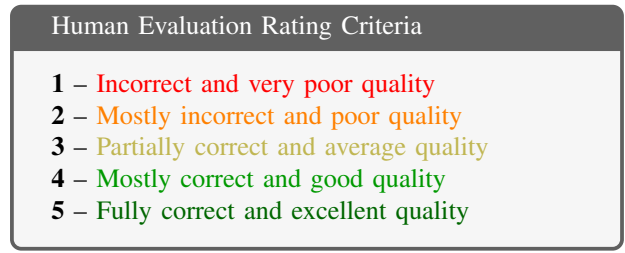

    \centering
    \begin{tcolorbox}[colback=gray!7!white, colframe=gray!75!black, title=Human Evaluation Rating Criteria, width=0.45\textwidth]
    \textbf{1} -- \textcolor{red}{Incorrect and very poor quality} \\
    \textbf{2} -- \textcolor{orange}{Mostly incorrect and poor quality} \\
    \textbf{3} -- \textcolor{yellow!70!black}{Partially correct and average quality} \\
    \textbf{4} -- \textcolor{green!60!black}{Mostly correct and good quality} \\
    \textbf{5} -- \textcolor{green!40!black}{Fully correct and excellent quality}
    \end{tcolorbox}
    \caption{Interpretation of human evaluation scores (1–5).}
    \label{rating_scale}
\end{figure}

Each evaluator rates the responses on a scale of 1 to 5, based on correctness and quality. Correctness is assessed by comparing the responses with the ground-truth answers, while quality is determined based on grammar and writing style. The interpretation of the evaluation scores is illustrated in Figure \ref{rating_scale}. The final score for each response is computed as the average of the individual scores assigned by the evaluators. Finally, the overall human evaluation score is determined by averaging the scores across all responses.

\begin{table*}[!t]
    \centering
    \small
    \setlength{\tabcolsep}{6pt} % Adjust column spacing
    \renewcommand{\arraystretch}{1.2} % Adjust row spacing
    \caption{Comparison of evaluation results on the test set using two metrics: the average human evaluation score (1–5) and mean cosine similarity ($\mu$) ± standard deviation ($\sigma$). These metrics capture both the central tendency and variability in semantic similarity. The Advanced RAG (Ours) achieves superior performance compared to the Vanilla RAG.}
    \label{main_result}
    \begin{tabular}{l c c c c c}
        \toprule
        \textbf{Models} & \textbf{Parameters} & \multicolumn{2}{c}{\textbf{Vanilla RAG}} & \multicolumn{2}{c}{\textbf{Advanced RAG (Ours)}} \\
        \cmidrule(rl){3-4} \cmidrule(rl){5-6}
        & & \textbf{Human Evaluation} & \textbf{Cosine Similarity} & \textbf{Human Evaluation} & \textbf{Cosine Similarity} \\
        & & Out of 5 & $\mu$ ± $\sigma$ & Out of 5 & $\mu$ ± $\sigma$ \\
        \midrule
        Mixtral 8x7B & 7B & 2.77 & $0.70 \pm 0.120$ & 3.09 & $0.76 \pm 0.117$\\
        Llama 3.1 & 8B & 3.41 & $0.76 \pm 0.114$ & 3.70 & $0.82 \pm 0.101$\\
        Gemma 2 & 9B & 3.02 & $0.74 \pm 0.124$ & 3.28 & $0.81 \pm 0.120$\\
        \bottomrule
    \end{tabular}
\end{table*}

\begin{table*}[!t]
    \centering
    \small
    \setlength{\tabcolsep}{5pt} % Adjust column spacing
    \renewcommand{\arraystretch}{1.2} % Adjust row spacing
    \caption{Domain-wise mean cosine similarity of the employed LLMs: Mixtral 8×7B, Llama 3.1 (8B), and Gemma 2 (9B). For all models, our proposed Advanced RAG method outperforms the Vanilla RAG pipeline. The \colorbox{gray!20}{gray-highlighted} rows refer to the results of our proposed approach, while the \textbf{bold} text indicates the best result for each domain.}
    \label{domain_wise}
    \begin{tabular}{l c c c c c c c c c}
        \toprule
        \multirow{2}{3em}{\textbf{Method}} & \multirow{2}{3em}{\textbf{Models}} & \textbf{Factual} & \textbf{Temporal} & \textbf{Gazette} & \textbf{Bangla} & \textbf{Statistical} & \textbf{Grammar/} & \textbf{Out of} & \multirow{2}{3em}{\textbf{Others}}\\
        
         & & \textbf{Question} & \textbf{Changes} & \textbf{Search} & \textbf{Dialect} & \textbf{Question} & \textbf{Spell Error} & \textbf{Context} & \\
        \midrule
         Vanilla RAG & Mixtral 8x7B & 0.77 & 0.68 & 0.71 & 0.69 & 0.69 & 0.72 & 0.41 & 0.77\\
         \rowcolor{gray!20} Advanced RAG (Ours) & Mixtral 8x7B & 0.77 & \textbf{0.76} & 0.74 & 0.70 & 0.77 & 0.79 & 0.44 & 0.78\\

         Vanilla RAG & Llama 3.1 (8B) & 0.80 & 0.73 & 0.80 & 0.69 & 0.68 & 0.77 & 0.45 & 0.79\\
         \rowcolor{gray!20} Advanced RAG (Ours) & Llama 3.1 (8B) & \textbf{0.87} & 0.74 & \textbf{0.84} & \textbf{0.74} & \textbf{0.80} & 0.79 & 0.45 & \textbf{0.80}\\
         
         Vanilla RAG & Gemma 2 (9B) & 0.79 & 0.72 & 0.75 & 0.72 & 0.66 & 0.77 & 0.55 & 0.77\\
         \rowcolor{gray!20} Advanced RAG (Ours) & Gemma 2 (9B) & 0.81 & 0.74 & 0.79 & 0.72 & 0.73 & \textbf{0.80} & \textbf{0.58} & 0.78\\
        \bottomrule
    \end{tabular}
\end{table*}

\subsection{Results and Analysis}
In this section, we present the results obtained using our proposed Advanced RAG pipeline for the Bangla question-answering task on legal government gazettes, specifically Bangladesh Police Gazettes. For comparison, we also report the performance of the Vanilla RAG framework on the same evaluation set. The evaluation results for the Q\&A task are presented in terms of human evaluation scores (ranging from 1 to 5) and cosine similarity (mean and standard deviation). Table \ref{main_result} summarizes the overall human evaluation scores and cosine similarity (mean and standard deviation) on the test set for the two RAG pipelines: Vanilla RAG and our proposed Advanced RAG. As observed, Advanced RAG achieves higher human evaluation scores and mean cosine similarity while also demonstrating lower standard deviation in cosine similarity. This clearly highlights the effectiveness of our proposed approach in handling complex, unstructured legal documents and generating accurate answers. Among the three employed LLMs, Llama 3.1 (8B) achieved the highest average human evaluation score and the highest mean cosine similarity, further emphasizing its strong performance in response generation.

Since the test set is diverse, featuring questions from multiple domains, we present the domain-wise mean cosine similarity in Table \ref{domain_wise}. As observed, the Advanced RAG method consistently outperforms the conventional Vanilla RAG pipeline across all domains, demonstrating its generalizability. Additionally, we find that RAG systems perform relatively well when answering factual questions or retrieving specific information from the gazettes. However, for out-of-context queries, they tend to generate incorrect responses with high confidence, as reflected in the lower mean cosine similarity scores. This highlights a key limitation of retrieval-augmented systems in handling ambiguous or contextually irrelevant queries.

\subsection{Ablation Study}
To evaluate the impact of key components in our system, we conduct ablation studies focusing on two critical aspects: (1) the effect of sampling temperature during response generation and (2) the influence of instruction prompt language. These additional experiments provide insights into how these factors affect the overall performance of the Advanced RAG pipeline.

\begin{figure}[!t]
    \centering
    \includegraphics[width=\columnwidth]{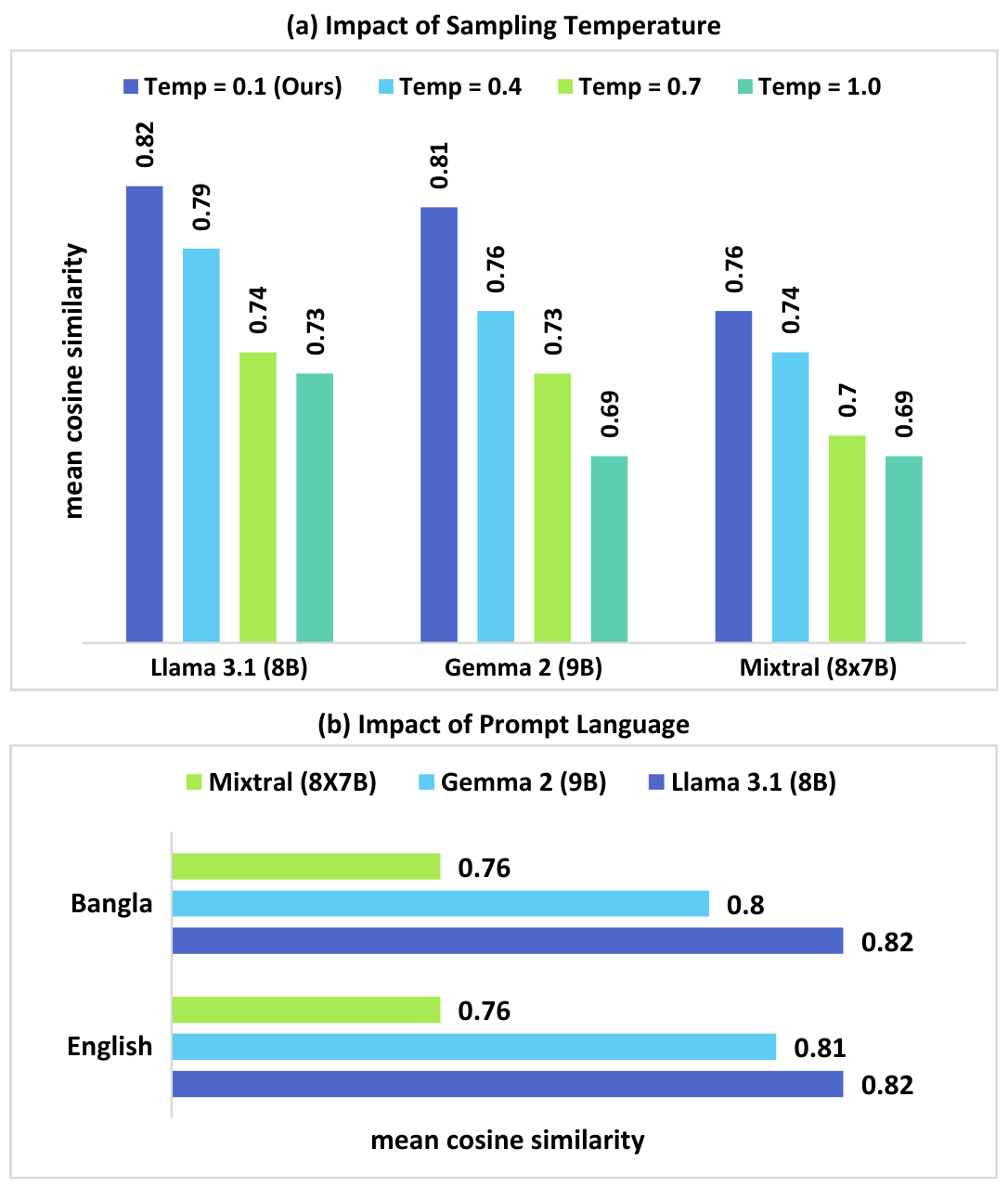}
    \caption{The impact of \textbf{(a)} sampling temperature and \textbf{(b)} prompt language on the responses generated by the Advanced RAG.}
    \label{ablation}
\end{figure}

\noindent\textbf{Impact of Sampling Temperature:}
The responses generated by large language models (LLMs) are significantly influenced by the sampling temperature. To analyze its impact on performance, we experimented with four distinct temperature values: 0.1, 0.4, 0.7, and 1.0, as shown in Figure \ref{ablation}(a). The figure reports the mean cosine similarities achieved by our proposed Advanced RAG framework for each sampling temperature. From Figure \ref{ablation}(a), we observe that all three LLMs achieve their highest mean cosine similarity when the temperature is set to 0.1. This aligns with the expectation that lower sampling temperatures facilitate more specific and precise responses, whereas higher values introduce greater randomness, which is beneficial for creative tasks. These findings are consistent with those of Renze \cite{renze-2024-effect}.

\noindent\textbf{Impact of Prompt Language:} 
In addition to sampling temperature, we also investigated the impact of the instruction prompt language on the system's performance. Since our source document consists of both Bangla and English texts, we conducted experiments by providing the instructions in these two languages, as shown in Figure \ref{ablation}(b). As observed, the mean cosine similarities achieved by the Advanced RAG framework for each prompt language are fairly similar. This can be attributed to the fact that our proposed pipeline utilizes a multilingual embedding model capable of understanding both Bangla and English texts. Therefore, the instruction prompt language does not significantly affect the system’s performance, suggesting that our approach is robust to multilingual inputs.

\subsection{Discussion}
Retrieving specific information from government regulatory documents is inherently challenging due to their complex structure and legal terminology \cite{ruhl2017harnessing, ruhl2019harnessing}, particularly in low-resource settings. This study presents a RAG-based approach for question-answering on Bangladeshi gazettes, addressing these challenges through an advanced retrieval mechanism.
Beyond the conventional RAG pipeline, we propose an enhanced multilingual RAG framework. The key innovation in our approach is the incorporation of an additional LLM that functions as both a relevance checker and a query refiner. This additional stage ensures that only the most appropriate text chunks are selected before being passed to the response generator LLM along with the user query. By refining the retrieval process, our advanced RAG pipeline significantly improves the accuracy and relevance of the generated responses.
The experimental results demonstrate that the proposed advanced RAG pipeline consistently outperforms the standard Vanilla RAG framework, highlighting its effectiveness in handling complex legal documents.

\section{Limitations and Future Works}
Table \ref{main_result} demonstrates that our proposed RAG pipeline achieves superior performance compared to the Vanilla RAG approach. However, significant limitations still remain. This is likely due to the insufficient exposure of LLMs to complex legal documents during their pre-training, which affects their ability to generate highly accurate responses in this domain. Additionally, as shown in Table \ref{domain_wise}, both the Vanilla RAG and our Advanced RAG pipeline struggle significantly with out-of-context questions. This highlights a key limitation of RAG-based systems in distinguishing relevant information when queries fall outside the scope of the documents. 

Future research should aim to address the identified limitations by fine-tuning LLMs on complex, low-resource legal documents to improve their familiarity with legal terminologies and domain-specific nuances. This fine-tuning process will require the creation of a comprehensive corpus of government-published documents to enhance the models' contextual understanding. Additionally, exploring advanced NLP techniques to improve the handling of out-of-context queries could significantly enhance the robustness and reliability of legal question-answering systems.

\section{Conclusion}
This study explores the application of computational linguistics to government regulatory documents \cite{frankenreiter2020computational}, an area that has received limited attention. By leveraging modern RAG-based techniques, we contribute to the advancement of legal information retrieval \cite{van2017concept, sansone2022legal, vsavelka2022legal}. Beyond conventional RAG systems, we introduce an advanced RAG pipeline that incorporates a relevance check and query refinement, ensuring more accurate text retrieval and improved question-answering performance. Facilitating efficient question-answering on government-published gazettes can significantly enhance access to legal and regulatory information. Our findings demonstrate that advanced NLP techniques like RAG can be effectively applied to low-resource legal documents, highlighting their potential for future development in legal AI applications \cite{zhong-etal-2020-nlp}.

\section*{Ethics Statement}

This study was conducted with a strong commitment to ethical standards, particularly given the sensitivity of handling legal documents. The data used, including Bangladesh Police Gazettes, was sourced from publicly available records, ensuring compliance with confidentiality and privacy requirements. No unpublished data was used at any stage of the research. We recruited three expert evaluators with proficiency in Bangla and English and extensive knowledge of Bangladeshi legal documents. Participation was entirely voluntary, and no personal or sensitive data were collected or stored. Measures were taken to safeguard the well-being of the evaluators, and the evaluation process posed no risks. System-generated responses were carefully reviewed to prevent bias or harmful content. By prioritizing ethical considerations in handling sensitive legal data, this research upholds ethical integrity.

\bibliographystyle{IEEEtran}
\bibliography{conf}

% \begin{thebibliography}{00}
% \end{thebibliography}

\end{document}